\documentstyle[preprint,aps]{revtex}
\newcommand{\be}{\begin{equation}}
\newcommand{\ee}{\end{equation}}
\newcommand{\ber}{\begin{eqnarray}}
\newcommand{\eer}{\end{eqnarray}}

\begin{document}
\draft

\title{Electron Counting Statistics and Coherent States of Electric Current}
\author{Leonid~S.~Levitov, Hyunwoo~Lee}
\address{Physics Department,
Massachusetts Institute of Technology,\\
12-112, 77 Massachusetts Ave., Cambridge, MA 02139}
\author{Gordey~B.~Lesovik}
\address{The Institute for Solid State Physics,
Chernogolovka, Moscow region, Russia}
\maketitle

\begin{abstract}
A theory of electron counting statistics in quantum transport is
presented. It involves an idealized scheme of current
measurement using a spin $1/2$ coupled to the current so that it
precesses at the rate proportional to the current. Within such
an approach, counting charge without breaking the circuit is
possible. As an application, we derive the counting statistics
in a single channel conductor at finite temperature and bias.
For a perfectly transmitting channel the counting distribution
is gaussian, both for zero-point fluctuations and at finite
temperature. At constant bias and low temperature the
distribution is binomial, i.e., it arises from Bernoulli
statistics. Another application considered is the noise due to
short current pulses that involve few electrons. We find the
time-dependence of the driving potential that produces coherent
noise-minimizing current pulses, and display analogies of such
current states with quantum-mechanical coherent states.
  \end{abstract}
\pacs{PACS numbers: 72.10.Bg, 73.50.Fq, 73.50.Td}
\narrowtext

\tableofcontents

\section{Introduction}
Quantum transport in nanostructures has been a subject of many
recent studies\cite{review}. Transport properties like Ohmic
conductivity can be understood in terms of the quantum
scattering problem in the conductor, which provides a theory of
quantum coherence of transport\cite{9}. Fluctuations of electric
current due to the discreteness of electric charge are intrinsic
to quantum transport\cite{2,3,4}. It has been found that current
fluctuations have interesting properties reflecting profound
aspects of underlying quantum dynamics\cite{5,6,7,deJong}.
For example, the quantum noise caused by a dc current is reduced
below classical shot noise level\cite{2,3,4,5,6,7,deJong}. This suppression
has been understood as an effect of enhanced regularity of
transmission events due to Fermi statistics\cite{LL}. Besides
theoretical interest, such phenomena may lead to applications.
Given the development of nano-technologies, the
transmission of signals by single-~ or
few-electron pulses will become common, and then one will see
the quantum statistics of current working.

In this paper we update the theory of quantum measurement of
electric current\cite{1'94}. Our goal is a complete description
of charge fluctuations, rather than developing measurement
theory (see Secs.~II,~III). We shall derive a microscopic
formula for electron counting distribution (see Sec.~III,
Eq.~(\ref{A9}), and Sec.~IV, Eq.~(\ref{A20})) that can be used
for any system, e.g., with an interaction or with a
time-dependent potential\cite{2'93}. As an application, we test
the method on the statistics in a single channel ideal conductor
for non-equilibrium and equilibrium noise at finite temperature,
and for zero-point equilibrium fluctuations (Secs.~IV,~V). In
particular, the fluctuations of a dc current at zero temperature
are found to be binomial (Sec.~VI) with the probabilities of
outcomes related with transmission coefficients of elastic
scattering in the system, and with the number of attempts
$N=eVt/h$, where $V$ is applied voltage, and $t$ is the time of
measurement.

Another property of quantum noise that does not have classical
analog is its phase sensitivity\cite{C9,2'94}. For the current
correlator $\langle\!\langle j(t_1)j(t_2)\rangle\!\rangle_+$ it
results in a periodic sinusoidal dependence on Faraday's flux
due to applied voltage, $\Phi=c\int^{t_2}_{t_1}V(t)dt$, with the
period $\Phi_0=hc/e$. The phase sensitivity manifests in
singularities of the low frequency noise power in a junction
driven by ac and dc signals together\cite{3'94}.

Even more remarkable is the behavior of current fluctuations
due to a pulse of voltage\cite{C9,3'95}. Total charge that flows through the
conductor due to a voltage pulse fluctuates in such a way that
the mean square fluctuation diverges whenever the flux of the pulse is
not an integer: $\varphi={e\over\hbar}\int_{-\infty}^{\infty}V(t)dt\ne
2\pi n$. On the other hand,
for $\varphi=2\pi n$ the fluctuation of the
transmitted charge is finite (Sec.~VII). This result has
simple interpretation in terms of the Anderson orthogonality
catastrophe theory, since the flux
$\varphi$ enters the time dependent scattering matrix of the
conductor through the forward scattering
amplitude.

With this, one is led to address the issue of current states
that minimize the current fluctuations at fixed mean transmitted
charge\cite{1'95,3'95}. It is found in Sec.~VIII
that such states are produced by time-dependent voltage of the form
  \be
V(t)=\ \pm{h\over \pi e}\
\sum_{k=1}^n{\tau_k\over(t-t_k)^2+\tau_k^2}\ , \tau_k>0\ ,
  \ee
a sum of Lorentzian pulses of unit flux each. It is
remarkable that the minimal noise due to such sequence of pulses
is independent of the pulse positions $t_k$ and widths
$\tau_k$, which leads to obvious parallels with solitons in the
theory of non-linear integrable systems. The noise minimizing
current states can be compared to the coherent states that
minimize the quantum-mechanical uncertainty. Apart from obvious
similarity, there is a difference: the coherent current states
are many-body time-dependent scattering states. Their role in
transport is an interesting subject of future work: one expects
that representing many-body states as a superposition of these
coherent states has an advantage similar to that provided by
coherent states of one particle.

\section{Measuring electric current}
Instantaneous measurement is described in quantum mechanics by
wavepacket reduction that involves projecting on eigenstates of
an observable, represented by a hermitian operator. A different
kind of measurement, extended in the time domain, is realized in
detectors and other counting devices. It is known that in such
cases a certain revision of the measurement description is
necessary. A famous example is the theory of photon
detectors\cite{MandelGlauber} in quantum optics. Due to Bose
statistics, photons entering a photo-counter are correlated in
time, and this makes the theory of photon detection a problem of
many-particle statistics. For a single normal mode of radiation
field the probability $P_m$ to count $m$ photons over time $t$
is given by
   \begin{equation}\label{A1}
P_m={(\eta t)^m \over m!}\langle\ :(a^+a)^m e^{-\eta t a^+a}:\ \rangle \ ,
\end{equation}
    where $a^+$ and $a$ are Bose operators of the mode, $\eta$
is the counter efficiency parameter, and $\langle ...\rangle$
stands for the average over a quantum state. The normal ordering
$:\dots:$ is an important element of the formalism. Physically,
it means that, after having been detected, each photon is
destroyed, e.g., it is absorbed in the detector. Instead of the
probabilities (\ref{A1}), it is more convenient to deal with the
generating function
  \begin{equation}\label{A2}\chi(\lambda)=\sum_m \ P_m e^{i\lambda m}\ .
\end{equation}
  For the single normal mode Eq.(\ref{A1}) leads to
  \begin{equation}\label{A3}\chi(\lambda)=
\langle\ :\exp \eta t (e^{i\lambda}-1) a^+a :\ \rangle \ .
\end{equation}
Eqs.(\ref{A1},\ref{A2},\ref{A3}) account very well for numerous
experimental situations
\cite{Gradinger}. Particularly interesting is the case of a
coherent state $|{\rm z}\rangle$, $a|{\rm z}\rangle=z\ |{\rm z}
\rangle$, where $z$ is a complex number. It corresponds to the
radiation field of an ideal laser, and with Eq.(\ref{A3}) one
easily gets Poisson counting distribution,
  \be
P_m={(Jt)^m \over m!}e^{-Jt},\qquad
J=\eta|z|^2\ ,
  \ee
which describes the so-called minimally bunched light source.

Similar to the photon detection, electric
measurement is performed on a system containing an enormous
number of particles --- in this case fermions --- and thus one
expects the effects of Fermi statistics to be important. Also,
the duration of electric measurement is typically much longer
than the time it takes the system to transmit one electron by
microscopic tunneling, scattering, or diffusion. Apart from these
similarities, there is, however, a crucial difference from the
photon counting: the number of electrons is not changed by the
current measurement, since electric charge is
conserved. This has to be contrasted with absorption of photons
in photo-counters. Related to this, there is another important
difference: at every detection of a photon, its energy
$\hbar\omega$ is taken from the radiation field, which makes
plain photodetectors insensitive to zero-point fluctuations of
electromagnetic field. On the contrary, the
measurement of current fluctuation is usually performed without
changing energy of the system, which makes the zero-point noise
an unavoidable component of any electric
measurement\cite{0-noise}. (Let us emphasize that the difference
has  nothing  to  do  with the type of quantum statistics, Fermi or Bose.
Rather it is the difference between the two kinds of
measurement, e.g., see \cite{e-count}, where counting of
fermions was discussed using an optical-like counter that has to
capture an electron in order to detect it.)

In the classical picture, the measurement gives the charge
$Q(t)=\int_0^t j(t')dt'$ transmitted during the measurement time
$t$. The probabilities $P_m$ of counting $m$ electrons can then
be obtained by averaging $\delta(Q(t)-me)$ over the state of the
system. In a quantum problem electric current is an operator,
and since currents at different moments do not commute, the
operator of transmitted charge $\widehat Q(t)=\int_0^t \widehat
j(t')dt'$ generally does not make any sense. Instead, since we
are interested in higher-order statistics of current
fluctuations, beyond $\langle \widehat j(t)\rangle$ and
$\langle\!\langle j(t_1)j(t_2)\rangle\!\rangle_+$,
in order to compute
electron counting ditribution, we have to include the measuring
system in the quantum Hamiltonian. Our approach is motivated by
the example of the quantum mechanical systems with strong
coupling to macroscopic environment, introduced by Leggett, that
can be treated consistently only by adding the ``measuring
environment'' to the quantum problem\cite{Leggett}.

For that we introduce a model quantum galvanometer, a spin $1/2$
that precesses in the magnetic field $B$ of the current. For a
classical system, the rate of precession is proportional to
$B(t)$, and $B(t)$ is proportional to the current $I(t)$:
$B(t)=const{\ }I(t)$. Therefore, the precession angle of the
spin directly measures transmitted charge
${\delta}Q={\int_0^t}I(t')dt'$. We adopt the same measurement
procedure for the quantum circuit, i.e., we include in the electron
Hamiltonian the vector potential due to the spin:
   \begin{equation}\label{A4}
\vec A(r)= -\mu\ \vec{\widehat\sigma}\times\vec\nabla{1\over|r|}\ ,
\end{equation}
  where
$\vec{\widehat\sigma}=(\sigma_x, \sigma_y, \sigma_z)$ are Pauli matrices.
Thus we obtain a Hamiltonian describing motion of electrons,
the measuring spin, and their coupling. Now, according to what
has been said, we have to solve dynamics of the spin in the
presence of the fluctuating current, find the distribution of
precession angles, and then interpret it as a distribution of
transmitted charge. Of course, a question remains about the back
effect of the spin on the system, as in any other problem of
quantum measurement. However, as we find below in (\ref{A15})
and (\ref{A14}),
only the phase of an electron state is affected by the presence
of the spin, not the amplitude. Moreover, the phase will change
only for the transmitted, but not for the reflected wave. As a
result, the probabilities we obtain do not depend on the
coupling constant of the spin. This justifies the assumption
that the spin measures charge transfer in a non-invasive way.

It is worth remarking that our scheme resembles the ``Larmor clock''
approach\cite{L-clock} to the problem of traversal time for motion through a
classically forbidden region.
In this problem one
is interested, e.g., in the time spent by a particle
tunneling through a barrier. The Larmor clock approach involves
an auxiliary constant magnetic field $B$ added in the
classically forbidden region, and a spin $1/2$ carried by the
particle that interacts with the field: ${\cal
H}_{int}=-\widehat\sigma_zB$. The precesson angle of the spin
measures traversal time. Comparing the two approaches is very tutorial:
see Appendix A, where the Larmor clock is reviewed.

\section{Spin $1/2$ as a galvanometer}
Having clarified our motivation, we proceed
semi-phenomenologically and choose a new vector
potential in the spin-current interaction $-{1\over c}\vec j
\vec A$. We replace the Amp\`ere's long-range form (\ref{A4}) by
a model vector potential
    \begin{equation}\label{A5}
\widehat A_i(r)= {\lambda\Phi_0\over 4\pi}\ \widehat\sigma_z\
\nabla_i\theta(f(r)-f_0)
\end{equation}
   concentrated on some surface $S$ defined by the equation
$f(r)=f_0$. Here $\Phi_0=hc/e$, $\lambda$ is a coupling
constant, $f(r)$ is an arbitrary function, and, as usual,
the step-function
$\theta(x)=1$ for $x>0$, $0$ for $x<0$.
The surface $S$ defines a section of the conductor on
which the interaction is localized:
    \be\label{A5a}
{\cal H}_{int}=
\int -{1\over c}\ \vec{\widehat j}\vec{\widehat A}\ d^3r=
-{\lambda \hbar\over 2e}\widehat\sigma_z\widehat I_S\ ,
   \ee
where $\widehat I_S= \int_S {\vec {\widehat j}}{\vec ds}$, i.e., the
spin now is coupled to the total current through the section
$S$. With the choice (\ref{A5}) of the vector potential one can
study current fluctuations in an arbitrary section of the
conductor. Another advantage of the phenomenological
Eq.(\ref{A5}) is that it involves only one Pauli matrix, which
makes the spin dynamics essentially trivial. The choice of the
quantization axis of the spin is arbitrary since (\ref{A5a})
will be the only spin-dependent part of the Hamiltonian.
Finally, another advantage of the form (\ref{A5a}) is that by
switching from the smooth function (\ref{A4}) to the singular
form (\ref{A5}) we enforce integer values of counted charge. To
understand this, let us note that in the ``fuzzy'' case
(\ref{A4}) the measurement can start at the moment when one of
the electrons is located somewhere in the middle of the volume
where $A\ne0$, and then a fractional part of electron charge
will be counted. On the contrary, in the ``sharp'' case
(\ref{A5}), the spin responds to the presence of an electron only
when it crosses the section $S$. We shall see below in a
microscopic calculation that integer values of charge follow
automatically from gauge invariance, since the form (\ref{A5})
is a gradient of a scalar.

Thus we come to the Hamiltonian
  \begin{equation}\label{A7}
\widehat{\cal H}_\sigma=\widehat{\cal H}(\tilde p, r),\qquad
\tilde p_i= p_i - {e\over c}\widehat A_i\ ,
\end{equation}
   where the spin-dependent $\vec{\widehat A}$ is taken in the  form
(\ref{A5}).  An  essential feature of our approach is that we treat the
constant $\lambda$ of coupling between the spin and the  current
as a {\it variable},  i.e.,  we consider the spin precession as a function of
the parameter $\lambda$. The reason is that, unlike  the  photon
counting  problem, our measurement scheme directly generates the
function $\chi(\lambda)$, and then  the  counting  probabilities
$P_m$ are obtained by reading Eq.(\ref{A2}) backwards.

At this point we are able to formulate our main result.  Let  us
define a new Hamiltonian
  \begin{equation}\label{A8}
\widehat{\cal H}_\lambda = \widehat{\cal H}(\tilde p, r),\qquad
\tilde p_i= p_i - {1\over 2}\lambda\hbar\ \nabla_i\theta(f(r)-f_0)\ ,
\end{equation}
    simply by supressing $\widehat\sigma_z$ in Eq.(\ref{A5}). The Hamiltonian
$\widehat{\cal H}_\lambda$ involves only quantities of the electron
subsystem. Below we show that by measuring precession of the spin
coupled to the current, one obtains the quantity
  \begin{equation}\label{A9}
\chi(\lambda)=\langle e^{i\widehat{\cal H}_{-\lambda}t}
e^{-i\widehat{\cal H}_{\lambda}t}\ \rangle\ .
\end{equation}
    Here  the  brackets  $\langle...\rangle$ stand for averaging
over initial state of electrons. Note  that  $\chi(\lambda)$  is
written  in  terms  of  a purely electron problem, not involving
spin variables. We shall find that the function  $\chi(\lambda)$
defines the result  of  any measurement of the spin polarization after
the time $t$ when the spin-current  coupling  is turned off.
Moreover,  we  shall  see that the function (\ref{A9}) has the meaning of
a generating function of electron counting distribution, i.e., the
Fourier   transform   of    $\chi(\lambda)$    gives    counting
probabilities, entirely analogous to (\ref{A2}).

Our  goal  now will be to express evolution of the spin in terms
of quantities corresponding to the electron system.
The interaction is given by Eqs.(\ref{A5}),(\ref{A7}).
Suppose that the measurement started at the moment $0$ and stopped at the
time $t$, i.e., the spin-current interaction is turned on
during the time interval $0<\tau<t$. Let us evaluate the density
matrix ${\widehat {\rho}}_s(t)$ of the spin, right after it is
disconnected from the circuit. We have
  \be
{\widehat  {\rho}}_s(t)={\rm tr}_e(e^{-i{\widehat{\cal  H}}_\sigma t} {\widehat
{\rho}}e^{i{\widehat{\cal H}}_\sigma t})\ ,
  \ee
where ${\widehat {\rho}}$ is initial density matrix ${\widehat
{\rho}}_e{\otimes}{\widehat {\rho}}_s$ at $t=0$, ${\widehat {\rho}}_e$
is initial density matrix of electrons,  and ${\rm tr}_e(...)$ means
partial trace taken over electron states, the spin indices left
free. In terms of the spin variables, the operator
$e^{-i{\widehat{\cal H}}_\sigma t}$ is a function only of
${\widehat\sigma}_z$, and hence it is diagonal in spin:
  ${\langle}{\uparrow}|e^{-i{\widehat{\cal H}}_\sigma t}|{\downarrow}{\rangle}=
{\langle}{\downarrow}|e^{-i{\widehat{\cal H}}_\sigma t}|{\uparrow}{\rangle}=0$.
  In other words, if initially the spin is in a pure state, up
or down, it will not precess. For ${\widehat {\rho}}_s(t)$ this
remark yields:
  \begin{equation}
\label{A10}
{\widehat {\rho}}_s(t)=\left[\matrix{
  {\widehat {\rho}}_{\uparrow \uparrow}(0) &
  {\chi}({\lambda}){\widehat {\rho}}_{\uparrow \downarrow}(0) \cr
  {\chi}(-{\lambda}){\widehat {\rho}}_{\downarrow \uparrow}(0) &
  {\widehat {\rho}}_{\downarrow \downarrow}(0)\cr }\right] .
  \end{equation}
Here $\chi(\lambda)={\rm tr}_e( e^{-i\widehat{\cal H}_{\lambda}t}
{\widehat {\rho}_e} e^{i\widehat{\cal H}_{-\lambda}t} )$, where $
e^{-i\widehat{\cal H}_{\lambda}t} $ is the evolution operator for
the problem (\ref{A8}). Now, after the spin degrees of freedom
are taken care of by (\ref{A10}), we are left with a purely
electron problem, that involves only electron degrees of
freedom, but not the spin. By using the cyclic property of the
trace ${\rm tr}_e(...)$, one can show that ${\chi}({\lambda})$
in Eq.(\ref{A10}) is identical to (\ref{A9}).

In principle, any entry of a density matrix can be measured, and therefore
the quantity $\chi(\lambda)$ is also measurable. In order
to make clear the relation of $\chi(\lambda)$ with the
distribution of precession angles, let us recall the
transformation rule for the spin $1/2$ density matrix under
rotation by an angle $\theta$ around the $z-$axis:
  \begin{equation}
\label{A11}
{\cal R}_\theta(\widehat\rho)=\left[\matrix{
  {\widehat {\rho}}_{\uparrow \uparrow} &
  e^{-i\theta}{\widehat {\rho}}_{\uparrow \downarrow} \cr
  e^{i\theta}{\widehat {\rho}}_{\downarrow \uparrow} &
  {\widehat {\rho}}_{\downarrow \downarrow}\cr }\right] .
  \end{equation}
By combining this with Eq.(\ref{A2})  we write $\widehat\rho_s(t)$ as
   \begin{equation}
\label{A12}
{\widehat\rho}_s(t)=\sum_m P_m{\cal R}_{\theta=m\lambda}(\widehat\rho)\ ,
\end{equation}
    which assigns to $P_m$ the meaning of the probability to observe
precession angle $m\lambda$. Let us finally note that such
interpretation of $P_m$ is consistent with what one expects on
classical grounds, because for a {\it classical} magnetic moment
$\vec\sigma$ interacting with the current according to (\ref{A5}), the angle
$\theta=\lambda$ corresponds to the precession due to a
current pulse carrying the charge of one electron.

\section{Single-channel conductor. General formalism}
In order to see Eq.(\ref{A9}) working,
let us consider an ideal single channel
conductor, i.e., the Schr\"odinger equation
   \begin{equation}\label{A13}
i{\partial\psi\over\partial t}=\left[{1\over
2}\left(-i{\partial\over\partial x }-{\lambda\over 2}\delta(x)\right)^2+U(x)
\right]\psi
\end{equation}
   in one dimension, where the potential $U(x)$ represents
scattering region and the vector potential is inserted according
to (\ref{A5}) at the $x=0$ section. In order to describe
transport, we shall use scattering states, left
and right. Their energy distributions $n_{L(R)}(E)$ are equilibrium Fermi
functions with temperature $T$ and chemical potentials shifted
by $eV$, $\mu_L-\mu_R=eV$, representing a dc voltage.

For the problem (\ref{A13}) one can  write  time  dependent
scattering states as
  \ber\label{scatt}
{\psi}_{L,k}(x,t)=e^{-iE_kt}\left\{\begin{array}{ll}
e^{ikx} +B_Le^{-ikx}  , & x<-a/2 \\
e^{i\lambda/2} A_Le^{ikx}, & x>a/2 \end{array}
\right.\cr
{\psi}_{R,k}(x,t)=e^{-iE_kt}\left\{\begin{array}{ll}
e^{-i\lambda/2}
A_Re^{-ikx}, & x<-a/2 \\
e^{-ikx}+B_Re^{ikx}, & x>a/2 \end{array}
\right.
  \eer
where $a$ is the width of the barrier, and $A_{L,R}$ and
$B_{L,R}$ are the transmission and reflection amplitudes in the
absence of the spin vector potential. To make expressions less
heavy, we supress electron spin. The phase factors $e^{\pm
i\lambda/2}$ in (\ref{scatt}) are found immediately by
observing that the vector potential in the Schr\"odinger
equation can be eliminated by the gauge transformation $\psi(x)
\rightarrow \exp(i\lambda / 2\ \theta(x)) \psi(x)$. The scattering
amplitudes form a unitary matrix:
  \begin{equation}
\label{A15}
\widehat S_\lambda =\left[\matrix{
{e^{i\lambda/2}A_L}&
B_R \cr
B_L&
{e^{-i\lambda/2}A_R}\cr }\right]
   \end{equation}
We will study the range of small $T,eV\ll E_F$, when only the
states close to the Fermi level are important. In this case,
there is an additional simplification because the states near Fermi
energy have almost linear dispersion, and thus all
wavepackets travel with the speed $v_F$ without changing shape.
Then, following Landauer and Martin\cite{7}, instead of the
usual scattering states (\ref{scatt}), it is convenient to use
their Fourier transform. By ignoring the energy-dependence of
$A_{L,R}$ and $B_{L,R}$, which is equivalent to saying that the
scattering time is negligible, and assuming that the dispersion
is stricktly linear, one obtains the representation of
scattering in terms of time-dependent scattering wave packets
    \ber \label{A14}
{\psi}_{L,\tau}(x,t)=
\left\{ \begin{array}{ll}
\delta(x_-),& t<\tau \\
e^{i\lambda/2}A_L\delta(x_-)+B_L\delta(x_+),& t>\tau
    \end{array}\right. \cr
{\psi}_{R,\tau}(x,t)=
\left\{ \begin{array}{ll}
\delta(x_+), & t<\tau \\
e^{-i\lambda/2}A_R\delta(x_+)+B_R\delta(x_-), & t>\tau
    \end{array}\right.
   \eer
  where $x_\pm=x\pm v_F(t-\tau)$. Here $\tau$ is the packet
arrival moment, at which it is scattered. It is straightforward
to verify orthogonality of the states (\ref{A14}). Any
time-dependent electron state can be written as a superposition
of the states (\ref{A14}), with the arrival time $\tau$ serving
in this representation as a label in the continuum of states,
like $k$ in (\ref{scatt}). The assumption that the scattering
amplitudes are energy-independent (and thus the scattering takes no
time) is equivalent to replacing the barrier $U(x)$ of finite
width by $U_0{\delta}(x)$ and is consistent with the closeness
of relevant energies to $E_F$.

Second-quantized, electron states (\ref{A14}) lead to
${\widehat {\psi}}(x,t)={\widehat {\psi}}_L(x,t)+{\widehat {\psi}}_R(x,t)$ with
   \begin{equation}
\label{A16}
{\widehat {\psi}}_{L(R)}(x,t)={\sum\limits_\tau}{\psi}_{L(R),\tau}(x,t)
{\widehat c}_{1(2),\tau}\ ,
\end{equation}
    where $c_{1,\tau}$ and $c_{2,\tau}$ are canonical Fermi
operators corresponding to the states (\ref{A14}), the left and
the right respectively. One checks that fermionic commutation
relations for $c_{1(2),\tau}$,
  \ber\label{anticommutators}
c^+_{i,\tau} c_{j,\tau'}+ c_{j,\tau'} c^+_{i,\tau}
&=&\delta_{ij}\delta(\tau-\tau') \\
c_{i,\tau} c_{j,\tau'}+ c_{j,\tau'} c_{i,\tau} =0\ &,&
c^+_{i,\tau} c^+_{j,\tau'}+ c^+_{j,\tau'} c^+_{i,\tau} =0\ ,
  \eer
yield the usual commutation relations for $\psi_{L(R)}(x,t)$.
From that one finds the meaning of the summation in (\ref{A16}):
$\sum\limits_\tau ...=\int^{\infty}_{-\infty}...d\tau$.
Mathematically, in this paragraph we defined second-quantized
${\psi}(x)$ in (\ref{A13}).

The advantage of introducing the basis of the wavepackets
(\ref{A14}),(\ref{A16}) is that now it is straightforward to
write the many-particle evolution operator through the
single-particle scattering matrix ${\widehat S}_\lambda$:
   \begin{equation}
\label{A17}
e^{-i{\widehat {\cal H}}_{\lambda}t}= \exp\int_0^t d\tau \sum\limits_{ij}
\ln[{\widehat S}_\lambda]_{ij} c_{i,\tau}^+c_{j,\tau} \ ,
   \end{equation}
where ${\widehat S}_\lambda$ is given by (\ref{A15}). To verify
(\ref{A17}), let us note that in the wavepacket representation
(\ref{A14}), according to Eqs.~(\ref{anticommutators}),
Fermi correlations occur only for the pairs of left and right
states that scatter at the same instant of time. For each of
such pairs the evolution operator $e^{-i{\widehat {\cal
H}}_{\lambda}t}$ is $\widehat 1$ if both states are occupied or both
are empty, otherwise it is given by the single-particle
scattering matrix (\ref{A15}).

Using similar arguments, we compute
   \begin{equation}
\label{A18}
e^{i{\widehat {\cal H}}_{-\lambda}t} e^{-i{\widehat {\cal H}}_{\lambda}t}=
\exp\int_0^t d\tau \sum\limits_{ij} {\widehat W}_{ij}
c_{i,\tau}^+c_{j,\tau}\ ,
\end{equation}
   where $e^{\widehat W}= {\widehat S}_{-\lambda}^{-1} {\widehat S}_\lambda$
is readily obtained from (\ref{A15}):
  \begin{equation}
\label{A19}
e^{\widehat W}=
\left[\matrix{
{e^{i\lambda}|A_L|^2+|B_L|^2}&
2i\sin\lambda\ \bar A_LB_R \cr
2i\sin\lambda\ \bar B_RA_L &
{e^{-i\lambda}|A_R|^2+|B_R|^2} \cr } \right]
\end{equation}
   Using unitarity of $e^{\widehat W}$ and commutation rules for
$c_{\alpha,\tau}$ one can rewrite (\ref{A18}) in terms of normal
ordering:
   \begin{equation}
\label{A20}
e^{i{\widehat {\cal H}}_{-\lambda}t} e^{-i{\widehat {\cal H}}_{\lambda}t}=\
:\exp\int\limits_0^t d\tau \sum\limits_{ij} [e^{\widehat W}-1]_{ij}
c_{i,\tau}^+c_{j,\tau}:
\end{equation}
   This form is ready to be plugged into Eq.(\ref{A9}) and averaged
over the initial state. Let us note the striking similarity of two
formulas obtained by different means: the fermionic Eq.(\ref{A20}) and the
bosonic Eq.(\ref{A3}).

Also, let us mention that the periodicity of the matrix
(\ref{A19}) in $\lambda$ ensures periodicity of $\chi(\lambda)$,
and thus guarantees integer values of charge.

\section{Equilibrium fluctuations}
Let us start with a simple one-particle
example. Consider
a particle in the
state $c^+_{1,\tau}|{\rm vac}\rangle$, which corresponds to
scattering at the moment $\tau$. In this case, from (\ref{A20})
and (\ref{A9}) one gets
  \be
\chi(\lambda)=\cases{
e^{i\lambda}|A|^2+|B|^2 & for $0<\tau<t$; \cr
1 &  otherwise; \cr }
  \ee
$|A|=|A_L|=|A_R|$,
$|B|=|B_L|=|B_R|$. Evidently, according to Eq.(\ref{A2}), this
simply means that for the scattering occurring during operation
of the detector, the counting probabilities are identical to the
one-particle scattering probabilities, as it should be expected.

Now, we consider current fluctuations in an equilibrium Fermi
gas. First, let us assume perfect transmission: $B_{L(R)}=0$.
Then Eq.(\ref{A19}) gives $\widehat W=i\lambda\sigma_z$, and thus
Eq.(\ref{A18}) becomes
   \begin{equation}
\label{A21}
e^{i{\widehat {\cal H}}_{-\lambda}t} e^{-i{\widehat {\cal H}}_{\lambda}t}=
\exp i\lambda
\int_0^t (c_{1,\tau}^+c_{1,\tau}- c_{2,\tau}^+c_{2,\tau}) d\tau \ ,
\end{equation}
   i.e., the right and the left states separate.

We observe that averaging of (\ref{A21}) over the Fermi ground state
is identical  to  that  performed  in the orthogonality catastrophe
calculation\cite{Mahan,ortho}. Hence, averaging of (\ref{A21}) can
be done by using the bosonization method\cite{Haldane} that
replaces the fermionic Hamiltonian by a bosonic one. (The
calculation is described in Appendix B.) In the bosonized
representation one has to do a simple gaussian average, which
gives
   \begin{equation}
\label{A22}
\chi(\lambda)= e^{-\tilde\lambda^2f(t,T)}\ ,
   \end{equation}
where $\tilde\lambda/2\pi+1/2=\lfloor\lambda/2\pi+1/2\rfloor$,
with $\lfloor...\rfloor$ being the fractional part. The function
  \ber\label{compute-f}
f(t,T)&=&
\left\langle\!\!\left\langle\left(\int_0^t c_{1,\tau}^+c_{1,\tau}d\tau\right)^2
\right\rangle\!\!\right\rangle
=-{\rm Re}\ {1\over4}\int\limits_0^t\!\int\limits_0^t
{T^2 dt_1 dt_2\over \sinh^2(\pi T(t_1-t_2-i\delta))} \\
&=&{1\over2\pi^2}\ln\Bigl({1\over\pi T\delta}\sinh\pi T t\Bigr)=
\left\{\begin{array}{lcr}
 {1\over2\pi^2}\ln t/\delta &,& \delta\ll t\ll\hbar/T \\
Tt/h-{1\over2\pi^2}\ln 2\pi T\delta &,& t\gg\hbar/T
\end{array}
\right.\ ,
   \eer
where $\delta$ is an ultraviolet cutoff time, of the order of $\hbar/E_F$.
At long times, according to (\ref{A2}), this leads to gaussian
counting statistics.

As a side remark, the distribution given by Eq.(\ref{A22}) also
gives a solution to another problem: the statistics of the
number of fermions inside a segment of fixed length in one
dimension. The relation is immediately obvious after one assigns
to $\tau$ in Eq.(\ref{A21}) the meaning of a coordinate on a
line. Thus, in this problem the statistics are gaussian as well.

Now, it turns out that the general case of non-vanishing
reflection, $B\ne0$, can be reduced to (\ref{A21}) by a
canonical Bogoliubov transformation of $c_{\alpha,\tau}$ making
the quadratic form in (\ref{A18}) diagonal. The transformation
is related in the usual way with the eigenvectors of the matrix
$\widehat W$. Thus, we come to Eqs.(\ref{A21}),(\ref{A22}) with
$\lambda$ replaced by $\lambda_*$:
  \be
\sin{\lambda_*\over2}=|A|\sin{\lambda\over2}\ .
  \ee
The counting statistics in this case are non-gaussian:
   \begin{equation}
\label{A23}
\chi(\lambda)= e^{-\lambda_*^2f(t,T)}\ .
   \end{equation}
One checks that the second moment of the distribution
   \be
\langle\!\langle m^2\rangle\!\rangle=
-\left.{\partial^2\chi(\lambda)\over\partial\lambda^2}\right|_{\lambda=0}=
2|A|^2f(t,T)
   \ee
agrees with the Johnson-Nyquist formula for the equilibrium noise.

\section{Statistics of a dc current: quantum shot noise}
Let us consider non-equilibrium noise.
In this case, due to the asymmetry in the distributions,
$n_{L(R)}(E)=(\exp (E{\pm}{1\over2}eV)/T+1)^{-1}$, generally one
cannot uncouple the two channels by a canonical transformation.
We calculate the statistics within an approximation that
ignores the effect of switching at $\tau=0$ and $\tau=t$.
Let us close the axis $\tau$ into a circle of
length $t$, which amounts to restricting on periodic states:
  \be
{\psi}(\tau)={\psi}(\tau\pm t)\ .
  \ee
For the $t-$periodic problem, by going to the Fourier space, one
has
  \ber
{\chi}(\lambda)&=& \prod\limits_{k\in Z}
\left[1+|A|^2(e^{-i\lambda}-1)n_L(E_k)(1-n_R(E_k))
\right.\cr
& &\left.+ |A|^2(e^{i{\lambda}}-1)n_R(E_k)(1-n_L(E_k))\right]\ ,
  \eer
where $E_k=2\pi\hbar k/t$, $k$ is an integer. For large $t$,
$t\gg\hbar/T$ or $t\gg\hbar/eV$, the product is converted to an
integral:
  \ber
\ln({\chi}&(&\lambda))=
{t\over 2{\pi}{\hbar}}{\int_{-\infty}^{+\infty}}
dE\ \ln\left(1+|A|^2 (e^{-i\lambda}-1)
\right.\cr
& &\left. \times
n_L(1-n_R)+|A|^2(e^{i{\lambda}}-1)n_R(1-n_L)\right)\ .
  \eer
We evaluate it analytically, and get
  \begin{equation} \label{A24}
{\chi}({\lambda})=
\exp\left(-tT u_+u_-/h\right)\ ,
  \end{equation}
where
  \be
u_\pm=v\pm\cosh^{-1}(|A|^2\cosh(v+i\lambda)+|B|^2\cosh v)\ ,
  \ee
$v=eV/2T$. The answer simplifies in the two limits: $T\gg eV$
and $eV\gg T$. In the first case we return to the equilibrium
result (\ref{A23}). In the second case, corresponding to the
recently discussed quantum shot noise\cite{2,3,4}, we have
  \begin{equation}
\label{A25}
{\chi}({\lambda})=(e^{i\epsilon\lambda}|A|^2+|B|^2)^{e|V|t/h},
\ \epsilon={\rm sgn}V\ ,
\end{equation}
   Analyzed according to Eq.(\ref{A2}), this
${\chi}({\lambda})$ leads to the binomial distribution
 \[P_N(m)=p^mq^{N-m}C_N^m\ ,\]
$p=|A|^2$, $q=|B|^2$, $N=e|V|t/h$. One checks that the moments
${\langle}m{\rangle}=pN$ and
$\langle\!\langle m^2\rangle\!\rangle=pqN$ correspond
directly to the Landauer formula and to the formula for the
intensity  of the quantum shot noise\cite{2,3,4}.
The correction to the statistics due to  the  switching  effects
is insignificant \cite{LL}.

\section{Noise due to a voltage pulse: Orthogonality catastrophe}
Here we consider the fluctuations of current in a single-channel
conductor induced by a voltage pulse. The result will be that
the dependence of the fluctuations on Faraday's flux $\Phi=
-c\int V(t)dt$ contains a logarithmically divergent term
periodic in $\Phi$ with the period $\Phi_0=hc/e$. The
fluctuation is smallest near $\Phi=n\Phi_0$. The divergence is
explained by a comparison with the orthogonality catastrophe
problem. The $\Phi_0-$periodicity is related with the
discreteness of ``attempts'' in the binomial statistics picture
of charge fluctuations presented above.

Initially, the orthogonality catastrophe problem emerged from
the observation that the ground state of a Fermi system with a
localized perturbation is orthogonal to the non-perturbed ground
state, no matter how weak the perturbation\cite{Anderson}.
Originally, the discussion was focused on the purely static
effect of Fermi correlations on the ground state that leads to
the orthogonality, but then it shifted to dynamical effects.
When a sudden localized perturbation is turned on in a Fermi
gas, the number of excited particle-hole pairs detected over a
large time interval $t$ diverges as $\ln t/\tau$, where $\tau$
is the time of switching of the perturbation. This effect leads
to power law singularities in transition rates
involving collective response of fermions, such as X-ray
absorption in metals\cite{Nozieres,Mahan}. In this section we
present an application of the orthogonality catastrophe picture
to the electric current noise.

Let us consider a single channel conductor in an external field
described by the one-dimensional Schr\"odinger equation
  \ber
& &i{\partial\over \partial t}\psi (x,t) = \widehat{\cal H} \psi (x,t)\ ,\cr
& &\widehat{\cal H} =
{1\over 2}\left(-i{\partial\over\partial x }-{e\over c}A(x,t)\right)^2+U(x)\ ,
   \eer
where the potential $U(x)$ represents the scattering region and
$A(x,t)$ is the vector potential corresponding to the applied
pulse of electric field. Since the pulse duration $\tau$ is
assumed to be much longer than the time of scattering, one can
treat the vector potential as static and apply a gauge
transformation in order to accumulate the flux $\varphi(t)=
e/\hbar\int^t_{-\infty}V(t')dt'$ in the phases of the
transmission amplitudes, thus making them time dependent. By
going through the argument presented in Sec.~IV, one obtains the
scattering states (\ref{scatt}) and (\ref{A14}) with
time-dependent forward scattering amplitudes:
   \begin{equation}\label{ampl}
A_{L(R)}\to A_{L(R)}\ e^{\pm i\varphi(t_r)}\ ,
   \end{equation}
where the time $t_r=t-|x|/v_F$ is taken retarded in order to account for
the finite speed of motion after scattering. As before, here  we
assume that scattering by the potential as well as traversing the region
where the voltage is applied takes negligible time  compared to
the duration of the voltage pulse. In this approximation the
amplitudes of backward scattering $B_{L(R)}$ are
time-independent constants.

To draw a relation with the orthogonality catastrophe problem,
let us study the effect of the voltage pulse on the scattering
phases $\delta_1$, $\delta_2$. They can be found by
diagonalizing the scattering matrix
  \begin{equation}\label{matr} \widehat{\cal S}(t)=\left[\matrix{
  A_Le^{i\varphi(t)+i\lambda/2} & B_R \cr
  B_L & A_Re^{-i\varphi(t)-i\lambda/2} \cr }\right]\ \ ,
  \end{equation}
and writing its eigenvalues as $e^{i\delta_1},e^{i\delta_2}$. The relation
between the phases $\delta_{1,2}$ before and after the pulse is
written conveniently through $\delta_\pm =(\delta_1\pm
\delta_2)/2$. The phase $\delta_+$ does not change at any time,
and the phase $\delta_-$ changes according to
   \be\label{deltaS}
\cos^2\delta_-(t')+\cos^2\delta_{-}(t)
-2\cos\delta_-(t')\cos\delta_-(t)
\cos\Delta\varphi =|A_L|^2\sin^2\Delta\varphi \ ,
   \ee
where $\Delta\varphi=\varphi(t')-\varphi(t)$.
Now, let us compare to the orthogonality catastrophe in the
Fermi system subjected to a time-dependent perturbation
(\ref{matr}). Change of the flux induces the shift of the phases
$\delta_\pm\rightarrow\delta_\pm'$ and makes the new ground
state orthogonal to the old one:
    \begin{equation}\label{orth} \langle 0' | 0\rangle =
\exp\left(-2{\delta^2_\ast \over\pi^2}\ln{L\over\lambda_F}\right)\ ,
\end{equation}
   where $L$ is the system size, $\lambda_F$ is Fermi wavelength, and
$e^{i\delta_\ast}$ is an eigenvalue of the matrix $\widehat{\cal
S}^{-1}(t=\infty)\widehat{\cal S}(t=-\infty)$:
  \be\label{deltaSS}
\sin{\delta_\ast\over2}=|A_L|\sin{\Delta\varphi\over2}\ .
  \ee
In terms of dynamics, this implies that the old ground state is
shaken up so that infinitely many particle-hole pairs are
excited\cite{Mahan}. It should lead to a logarithmically
diverging contribution to noise, since for each of the
particle-hole pairs there is a finite probability (equal to
$|A_LB_R|^2$) that the particle and the hole will go to
different terminals of the conductor, thus resulting in a
current fluctuation. The periodicity in Faraday's flux
$\Phi=-c\int V(t)dt$ follows from the gauge invariance and is
explicit in Eqs.(\ref{deltaS},\ref{deltaSS}) for $\delta_\pm'$. The
logarithmic divergence vanishes at $\Phi=n\Phi_0$, as
expected, since at integer $\Phi$ there is no long-term change
of the scattering.

Let us calculate the mean square fluctuation of the charge
$\langle\!\langle Q^2\rangle\!\rangle$ transmitted through the
system due to the pulse. For that, one can use the formula
(\ref{A20}) with the time-dependent scattering matrix
(\ref{matr}). To get the second cummulant $\langle\!\langle
Q^2\rangle\!\rangle$ one expands the exponent (\ref{A20}) up to
second order terms in $\lambda$, and takes an irreducible
average using Wick theorem.
The averages of $c_{i,\tau}$ have the usual form:
  \ber
& &\langle c^+_{i,\tau} c_{j,\tau'}\rangle=\delta_{ij}
\int n(E)e^{iE(\tau-\tau')}{dE\over2\pi}\ ,\cr
& &\langle c_{i,\tau} c^+_{j,\tau'}\rangle=\delta_{ij}
\int(1-n(E))e^{-iE(\tau-\tau')}{dE\over2\pi}\ ,
  \eer
where $n(E)= (e^{E/T}+1)^{-1}$ is the Fermi distribution.
The result reads
    \ber\label{1}
\langle\!\langle Q^2\rangle\!\rangle&=&
{{\rm g}e^2\over2\pi}\int\left(
|A|^4 {\Big |}\int_{0}^{t}e^{i\omega t'}dt'{\Big |}^2 + |AB|^2
\right.\cr
& &\left.
\times{\Big |} \int_{0}^{t}e^{i\varphi(t')+i\omega t'}dt'{\Big |}^2
\right) \omega{\rm coth}{\hbar\omega\over 2T}\ {d\omega\over
2\pi}\ ,
  \eer
where g is spin degeneracy.
The first term in (\ref{1}) is a part of equilibrium noise since
it does not depend on $\varphi$. To analyze the second term, let us
take a step-like time dependence of $\varphi$ resulting from an
abrupt voltage pulse applied at the time $t_0$, $0<t_0<t$, the
pulse duration $\tau$ being much shorter than $t$. Taking the
integral and keeping only the terms diverging at $t\to\infty$,
we find
   \ber\label{2}
{{\rm g}e^2\over2\pi} \int {\Big
|}{e^{i\omega t_0}-1\over i\omega} +e^{2\pi
i\Phi/\Phi_0}{e^{i\omega t}- e^{i\omega t_0}
\over i\omega}{\Big |}^2 |\omega|{d\omega\over2\pi}\cr
= {{\rm g}e^2\over\pi^2}
\left(
\ln {tE_F\over\hbar} +2\sin^2{\pi\Phi\over\Phi_0}\ln {t\over\tau}
\right)\ ,
  \eer
where the ultraviolet-diveregent integrals are cut at frequency
$\sim E_F/\hbar$. By subtracting the result for $\Phi=0$ as
corresponding to equilibrium, one obtains a logarithmic
contribution to the non-equilibrium noise:
   \begin{equation}\label{noise_result}
\langle\!\langle Q^2\rangle\!\rangle= {\rm g} e^2
|AB|^2 \Bigl[{2\over\pi^2}\sin^2{\pi\Phi\over\Phi_0} \ln{t_0\over\tau} +
{\Phi\over\Phi_0}\Bigr] +\dots+\langle\!\langle Q\rangle\!\rangle_{eq}\ ,
   \end{equation}
The origin of the non-diverging
term in Eq.~(\ref{noise_result}) proportional to $\Phi/\Phi_0$
will be discussed below.
The dots in Eq.~(\ref{noise_result}) represent corrections
higher order in $\Phi_0/\Phi$, and the equilibrium noise
   \be\label{equil}
\langle\!\langle Q^2\rangle\!\rangle_{eq}=
{e^2 G\over\pi^2} \ln{tE_F\over\hbar}\ ,\
G={\rm g}{e^2\over\hbar}|A|^2\ ,
   \ee
is obtained by repeating the calculation for $\Phi=0$. The expression
(\ref{equil}) agrees with the Nyquist formula
  \be
\langle\!\langle j_\omega j_{-\omega}\rangle\!\rangle= e^2 G\
\omega\ {\rm coth}{\omega\over 2T}
  \ee
taken at $T=0$, Fourier transformed, and combined
with the relation $Q=\int_{0}^{t}j(t')dt'$.

The term in Eq.~(\ref{noise_result}) proportional to $\Phi/\Phi_0$ is
obtained by rewriting the integral in the second term of (\ref{1}) as
    \begin{equation}\label{3}
\int\!\int\!\int  {d\omega\over 2\pi}
|\omega| dt_1 dt_2 e^{i(\varphi(t_1)-\varphi(t_2)+\omega(t_1-t_2))},
  \end{equation}
and extracting the contribution of almost coinciding times
$t_1$ and $t_2$ by going to new variables $t=(t_1+t_2)/2$,
$t'=t_1-t_2$, and changing the order of integrations:
   \begin{equation}\label{4}
\int dt\int {d\omega\over 2\pi}
|\omega|\int dt' e^{i\varphi(t_1)-i\varphi(t_2)+i\omega t'}=\
\int|{\dot\varphi}|dt \ ,
  \end{equation}
where we replaced
$\varphi(t_1)-\varphi(t_2)=
\varphi(t+t'/2)-\varphi(t-t'/2)$ by ${\dot\varphi}\
t'$. The result (\ref{4}) is approximate: it does not give the
log-term because the transformation (\ref{4}) properly takes
care of the integral (\ref{3}) only in the domain $t_1\simeq
t_2$, under the restriction that $\Phi(t)$ is varying
sufficiently smoothly. When $\Phi(t)$ is a monotonous function,
${\dot\varphi}>0$, the integral in the right hand side of (\ref{4}) equals
$2\pi\Phi/\Phi_0$ and thus produces the term of
Eq.~(\ref{noise_result}) proportional to $\Phi/\Phi_0$.

It is clear from the derivation that the two terms in the
brackets in
Eq.~(\ref{noise_result}) arise from different integration domains in
the $t_1$-$t_2$ space: the first term corresponds to
$|t_{1,2}|\ge\tau, \ t_1t_2<0$, while the second one is due to
almost coinciding moments, $|t_1-t_2|\ll\tau$. Since these domains
are almost non-overlapping, the two contributions to the noise
(\ref{noise_result}) do not interfere (cross terms are small).

In order to estimate the correction to the result
(\ref{noise_result}), let us derive it by another method
that allows to trace out the higher order
terms. For that, let us take the flux in the form
$\varphi(t)=N\lambda(t)$, where $\lambda(t)$ is a
smooth monotonous function, $\lambda(-\infty)=0$,
$\lambda(\infty)=2\pi$. For integer $N\gg 1$ the
Fourier component of $e^{iN\lambda(t)}$ entering Eq.~(\ref{4}) in
the stationary phase approximation is given by
    \be\label{B16} \int\limits_{-\infty}^\infty e^{iN\lambda(t)+i\omega t} dt =
\sum_k\ \sqrt{2\pi i\over N\ddot\lambda(t_k)} e^{iN\lambda(t_k)+i\omega
t_k} +...
   \ee
where the dots indicate terms $\sim {\rm O}(N^{-3/2})$, and
$t_k$'s   are   real   solutions   of   the   equation
$N\dot\lambda(t)+ \omega = 0$. Then we can write
    \be\left|
\int\limits_{-\infty}^\infty e^{iN\lambda(t)+i\omega t} dt\right|^2=
\sum_k\ {2\pi \over N\ddot\lambda(t_k)} \ +\ {\rm O}(N^{-2}) \ ,\ \ee
and thus obtain
    \be\label{B17}\langle\!\langle Q^2\rangle\!\rangle=\ A\
\int\limits_{-\infty}^\infty \sum_k\ { |\omega| d\omega
\over N\ddot\lambda(t_k)}\ +\ ... \ ,  \ee
  where the dots represent higher order terms. By differentiating
both sides of the equation $N\dot\lambda(t)= -\omega$ one finds the
relation $d\omega=-N\ddot\lambda(t_k)dt_k$, which means that
$|\omega| d\omega /\ddot\lambda(t_k) = - |\dot\lambda(t_k)| dt_k$, and
therefore the integral in Eq.~(\ref{B17}) equals
$N\int_{-\infty}^\infty d\lambda=2\pi N$. Since
$|\omega|d\omega$ scales as $N^2$, the correction to
Eq.~(\ref{B17}) can be evaluated as ${\rm O}(1)$, i.e., it is of
the order of one for any $N$. This means that Eq.~(\ref{B15}) has
relative accuracy of ${\rm O}(1/N)$.

The term in (\ref{noise_result}) proportional to $\Phi/\Phi_0$ is interesting in
connection with the picture of binomial statistics presented in Sec.VI.
In the dc bias case, the
distribution of charge for a single channel situation was found
to be binomial with frequency of attempts equal to $eV/h$ and
the probabilities of outcomes $p=|A|^2$, $q=|B|^2$.
Taken literally, this means that the attempts to transfer
charge are repeated regularly in time, almost periodic with the
period $h/eV$, with each attempt having two outcomes --
transmission or reflection -- occurring with the probabilities
$p$ and $q$. However, the regularity of the attempts does not
lead to an ac component in the current, rather it appears just as a
part of statistical description of charge fluctuations. Still,
the presence of a non-zero frequency in a non-interacting system
requires interpretation.

Let us suppose that the flux varies linearly with time,
$\Phi(t)=-cVt$. Since the ${\rm e.m.f.}=
-\partial\Phi/c\partial t$, the linear dependence
of $\Phi(t)$ is equivalent in its effect on the noise to
constant voltage $V$. In accordance with one's expectation,
the second term in the brackets in Eq.~(\ref{noise_result}) for a single channel is
$\langle\!\langle Q^2\rangle\!\rangle=\ {\rm g} e^2|AB|^2
\Phi/\Phi_0$, i.e., it is precisely of the form arising from the
binomial distribution with probabilities of outcomes $p$ and
$q$, and the number of attempts $N=\Phi/\Phi_0$. (Let us
recall that the second moment of the binomial distribution equals
$pqN$.) Taking into account that the time during which the flux
changes by $\Phi_0$ is $h/eV$, we can interpret the number of
attempts in the statistical picture as the number of flux quanta
by which the flux is changed. Such a conclusion suggests an
interesting generalization of the picture of binomial statistics
by attributing the meaning of the number of attempts to the flux
change measured in the units of $\Phi_0$, regardless of the linear or
non-linear character of the flux dependence on time.

It is appealing to put the special role of integer fluxes in
connection with the binomial statistics of current, where the
flux quanta are naturally interpreted as discrete attempts to
transmit charge. Although this picture is yet to be confirmed by
analytic treatment, it receives some support from the property
of the $\Phi_0-$periodic term in (\ref{noise_result}) to vanish
at every integer $\Phi$. One may conjecture that the statistics are
close to binomial only when the flux change is an integer, and
have diverging logarithmic corrections otherwise. The distinction
that Eq.~(\ref{noise_result}) makes between integer and
non-integer values of the flux and the relation of integer flux
change to the number of attempts in the binomial distribution,
gives another perspective to the statistical picture of a current pulse.

To summarize, the fluctuations caused by a voltage pulse, in
contrast to the average transmitted charge, distinguish between
integer and non-integer flux change. As a result, the dependence
of noise on the flux is non-monotonous and has minima at
integer values of the flux.

\section{Coherent states of current}
The question we address in this section is about  optimal  way  of
changing flux that minimizes induced noise.  It  is  clear  from
what  has  been said that to achieve minimum of the noise one
should change the flux by an integer amount,
    \be\label{B1}\Delta\varphi=
\varphi(t=\infty)-\varphi(t=-\infty)=2\pi n ,
    \ee
in order to suppress the logarithmically divergent
term. However, since for  a
given  $\Delta\varphi$  the  noise  depends  on the actual function
$\varphi(t)$, not just  on  $\Delta\varphi$,  we  have  a  variational
problem  to  solve  for  the  noise  as a functional of the time
dependence of the flux. This functional was derived in Sec.~VII.
At zero temperature it is given by
    \be\label{B2}\langle\!\langle Q^2\rangle\!\rangle=\ {{\rm g} e^2
\over2\pi} |AB|^2 \int\ {\Big |}
\int e^{i\varphi(t)+i\omega t} dt {\Big |}^2
|\omega| {d\omega\over 2\pi} \ ,  \ee
   where $A$ and $B$ are transmission and reflection
amplitudes, and ${\rm g}$ is spin degeneracy. We shall
study the variational problem (\ref{2}) with the boundary condition
(\ref{B1}) and show that its general
solution has the form of a sum of soliton-like functions:
    \be\label{B3}\Phi(t)=\pm{\Phi_0\over \pi} \sum_{k=1}^n \tan^{-1} \Bigl(
{t-t_k\over\tau_k} \Bigr) \ , \ \tau_k>0,  \ee
  where $t_k$ and $\tau_k$ are arbitrary constants. Under the
condition (\ref{B1}), any time dependence of the form (\ref{B3}) gives absolute
minimum to the noise:
    \be\label{B4}{\rm min}[\ \langle\!\langle Q^2\rangle\!\rangle\ ] =\
{\rm g} e^2 |AB|^2\ |n| \ .  \ee
  For an optimal time dependence of the  voltage  $V=-\partial\Phi/c\partial t$,
therefore, one  has  a  sum  of  Lorentzian peaks:
  \be V(t)=\ \mp{\Phi_0\over c\pi}\ \sum_{k=1}^n{\tau_k\over(t-t_k)^2+\tau_k^2}\ .
\ee
  In order to compare quantum noise  with  conductance,  let  us
mention  that the average  transmitted  charge
  \be
\langle\!\langle Q \rangle\!\rangle = {\rm  g}  e|A|^2{\Delta\varphi\over2\pi}
={\rm g}{e^2\over h}|A|^2\int V(t)dt
  \ee
simply obeys the Ohm's law, i.e., there is no particular
dependence on the way the flux change $\Delta\varphi$ is realized.

The result (\ref{B3}),(\ref{B4}) has a simple interpretation in
terms of the binomial statistics picture of charge fluctuations.
For the binomial distribution with probabilities of outcomes $p$
and $q$, $p+q=1$, and with the number of attempts $N$, the second
moment is known to be equal to $pqN$. The comparison with
Eq.~(\ref{B4}) suggests to attribute to $n=\Delta\Phi/\Phi_0$
the meaning of the {\it number of attempts}. This interpretation is
supported by the structure of the function (\ref{B3}) consisting
of $n$ terms, each corresponding to unit change of flux. A
remarkable property of the function (\ref{B3}) is its
separability, manifest both in the form of the terms and in the
way the parameters $t_k,\tau_k$ enter the expression. Let us
note that by making some of the $t_k$'s close to each other one
can have an overlap in time of different ``attempts''. The overlap,
however, does not change the fluctuations (\ref{B4}). The
situation reminds the one with solitons in integrable non-linear
systems, or with non-interacting instantons in integrable field
theories. Also, the absence of interference is interesting in
the context of coherent nature of transport in this system:
after all, we simply have scattering by a
time-dependent potential. Perhaps, proper interpretation of this
effect should be sought in establishing relation with the theory
of coherent states, known to eliminate to some
extent the quantum mechanical interference.

Let us now turn to the variational problem. It is convenient to
do the integral over $\omega$ first and to rewrite (\ref{2}) as
     \be\label{B5}\langle\!\langle Q^2\rangle\!\rangle=\ -{D\over\pi}\ \int\int
{e^{i\varphi(t)-i\varphi(t')}\over(t-t')^2}dtdt'\ ,\ee
  where $D= {{\rm g}e^2\over2\pi}|AB|^2$. In order to avoid
divergence at $t=t'$ the denominator in (\ref{B5}) should be understood
as
    \be\label{B6}{1\over 2}{\Big [}{1\over(t-t'+i\delta)^2}+
{1\over(t-t'-i\delta)^2}{\Big ]}\ ,\ \ \delta\to 0\ ,
   \ee
the condition that one obtains by introducing
regularization in (\ref{2}): $|\omega| \rightarrow |\omega|
e^{-|\omega|\delta}$. By considering variation of the
functional (\ref{B5}) we have the equation for an extremum:
     \be\label{B7}{\rm Im} {\Big [} e^{i\varphi(t)} \int {e^{-i\varphi(t')}
\over(t-t')^2}dt {\Big ]}=\ 0\ . \ee
   By using Cauchy formula one checks that the functions
  \be\label{B8}e^{i\varphi(t)}=\prod_{k=1}^n{t-\lambda_k\over t-{\bar\lambda}_k}
\ ,\ \lambda_k=t_k+i\tau_k\ , \ee
  satisfy (\ref{B7}) provided that $\tau_k$'s are all of the same sign. Obviously,
the functions (\ref{B8}) are just another form of (\ref{B3}).

It remains to be shown that the functional reaches its minimum
on the solutions (\ref{B8}). To prove it we proceed in the following
steps. Let us write $e^{i\varphi(t)}$ as
    \be\label{B9}e^{i\varphi(t)}=f_+(t)+f_-(t)\ ,   \ee
    where $f_+(t)$ and $f_-(t)$ are bounded  analytic  functions
in the upper and lower complex $t$ half-plane, respectively.
Representation (\ref{B9}) exists  for  any  non-singular  function  and
defines  the functions $f_+$ and $f_-$ up to a constant. Then we
substitute Eq.~(\ref{B9}) in (\ref{B5}), and  apply  Cauchy  formula  for  the
derivative,
    \be\dot f_\pm(t)=\pm{i\over2\pi}\oint{f_\pm(t')dt'\over(t-t'\pm i0)^2}\ , \ee
where  the  contour of integration is chosen in the half-plane of
analyticity of $f_+$ or $f_-$, respectively. Thus one gets
    \be\label{B10}\langle\!\langle Q^2\rangle\!\rangle=\ - i D\ \int
({\bar f}_+{\dot f}_+ - {\bar f}_-{\dot f}_-) dt\ . \ee
   On the other hand,
   \ber\label{B11}
n&=& {1\over 2\pi i} \int e^{-i\varphi(t)} {d \over dt}
e^{i\varphi(t)}dt \cr
&=& -{i\over 2\pi}\int ({\bar f}_+{\dot f}_+ +
{\bar f}_-{\dot f}_-) dt\ ,
   \eer
where the last equality is a result of substituting (\ref{B9}) and using
the relations
  \be
\int \bar f_+\dot f_- =\int \bar f_-\dot f_+=0\ ,
  \ee
which follow from Cauchy theorem. Now, Eq.~(\ref{B10}) can be rewritten through Fourier
components of $f_+$ and $f_-$ as
    \be\langle\!\langle Q^2\rangle\!\rangle=\ D\
\int\limits_0^\infty\ (|f_+(\omega)|^2+|f_-(-\omega)|^2)
\ \omega\ {d\omega\over2\pi} \ ,\ee
  thus demonstrating positivity of both terms in  (\ref{B10}).  (It  is
used  that  $f_+(\omega)=f_-(-\omega)=0$  for  $\omega<0$.) With
this, by comparing (\ref{B10}) and (\ref{B11}) we obtain
  \be\label{B12}
\langle\!\langle Q^2\rangle\!\rangle\ \ge\ 2\pi D\ |n|\ .
  \ee
Equality in (\ref{B12}) is reached only when either $f_+(t)$ or $f_-(t)$
vanishes. Therefore, to obtain the minimum one has to  take  the
functions  $e^{i\varphi(t)}$  that  are  regular in one of the
half-planes. This remark is sufficient to see that the functions  (\ref{B8})
form a complete family of solutions.

It is worth mentioning that the method used to derive (\ref{B12}) copies
almost entirely  the procedure of  derivation  of  the  duality
condition in the theory of instantons. Like in other  situations
where   the  duality  condition  holds,  our  ``solitons''  do  not
interact:  $\langle\!\langle   Q^2\rangle\!\rangle$   shows   no
dependence  on  the  parameters $\lambda_k$ of the solution (\ref{B8}).
Among numerous field theories that allow for exact  solution  of
the  instanton  problem  the  one  most  similar to our case is the
theory of classical Heisenberg ferromagnet in two dimensions.
For  this problem the  instantons  were  found by mapping the order
parameter space (i.e., the unit  sphere)  on  the  complex  plane\cite{C10}.
The duality  condition  was  shown  to take the form of the  constraint of
analyticity or anti-analyticity of the  mapped  order  parameter
function (compare with the condition $f_+=0$ or $f_-=0$ derived above).
Multi-instanton  solutions  were  given  as  products  of single
instanton  solutions  (cf.  Eq.~(\ref{B8})).  This  analogy  obviously
deserves more attention.

At this point let us examin an interesting non-optimal time
dependence of the flux, the sum of two solitons with opposite
charge:
    \be\label{B13}\varphi(t)=2 \Bigl[ \tan^{-1} \Bigl(
{t-t_1\over\tau_1} \Bigr) - \tan^{-1} \Bigl(
{t-t_2\over\tau_2} \Bigr) \Bigr] \ , \
  \ee
$\tau_{1,2}>0$. This function corresponds to $e^{i\varphi(t)}$ of
the form (\ref{B8}) but with the poles in both half-planes. In this
case $\Delta\varphi=0$, and thus $\langle\!\langle
Q\rangle\!\rangle=0$, so ${\rm min}[\ \langle\!\langle
Q^2\rangle\!\rangle\ ] =0$. With the function (\ref{B13}), however, one
finds
    \be\label{B14}\langle\!\langle Q^2\rangle\!\rangle=\ 4\pi D\ \Bigl|
{\lambda_1-\lambda_2\over \lambda_1-\bar\lambda_2} \Bigr|^2\
, \ee
  where   $\lambda_{1,2}=t_{1,2}+i\tau_{1,2}$.   For   different
values  of  the  parameters  $t_{1,2}$,  $\tau_{1,2}$  Eq.~(\ref{B14})
interpolates   between   two   trivial   limiting   cases:\\
   (i)
$\langle\!\langle  Q^2\rangle\!\rangle\to0$,  when  the  two flux
steps in (\ref{B13}) have nearly the same duration  and  almost  overlap;\\
  (ii)
$\langle\!\langle Q^2\rangle\!\rangle\to 4\pi A$, when the
flux steps either differ strongly in their duration  or  do  not
overlap.\\
  In the case (ii) the noise is twice bigger than the noise due to a
single step, as it should be.

We see that when $\Delta\varphi/2\pi$ is of the order  of  one  a
non-optimal  time  dependence $\varphi(t)$ can considerably enhance
the  noise.  It is not the case, however, for $\Delta\varphi/2\pi
\gg 1$. This limit was studied in Sec.~VII, where it was
found that when $\varphi(t)$ is a monotonous function the result
    \be
\label{B15}\langle\!\langle Q^2\rangle\!\rangle=\ {\rm g}e^2|AB|^2
|\Delta\varphi/2\pi|
\ee
   is rather accurate, even if the time dependence $\varphi(t)$ is not optimal\cite{C9}.

A more intuitive way to understand the accuracy of Eq.~(\ref{B15}) is to note
that for a given $n$ the number of parameters in the optimal
flux dependence (\ref{3}) is $2n$, which means that half of them are in some sense
redundant.
Because of that any smooth monotonous function with sufficiently
large variation $\Delta\varphi$ can be rather accurately  approximated  by  a
function  of  the  form (\ref{3}), and therefore the noise exceeds the
lower  bound  just  slightly.

An  implication  of  this  result  for  the  binomial statistics
picture is as follows. As it was  discussed  above  there  is  a
(conjectured) correspondence of the terms of Eq.~(\ref{B3}) and of the
attempts.  The deviation from the binomial distribution, that of
course  should exist for a non-optimal flux function $\varphi(t)$,
will remain bounded
in the case
of a smooth $\varphi(t)$,
as  $\Delta\varphi$
increases taking integer values. More precisely, the distribution
will  be  written  as  a mixture of binomial distributions with different
numbers $N$ of  attempts,  $P(m)=\sum_N\rho_NP_N(m)$, where $P_N(m)=p^mq^{N-m}C_N^m$. The estimated correction implies
that the distribution of attempts $\rho_N$ has finite variance
in the limit $N=\Delta\varphi/2\pi\to\infty$.

Before closing, let us mention that in order to apply the results
of Secs. VII, VIII to transport in a mesoscopic metallic
conductor with disorder, described by many conducting
channels with transmission constants $T_n$, one just needs to
replace $|AB|^2$ by $\sum_nT_n(1-T_n)$, since different
scattering channels contribute to the noise independently. The
condition of validity of our treatment then is that the
variation of the flux is sufficiently slow, so that ${\rm min}
[\tau_k ] \gg \hbar/E_c$, the time of diffusion across the
sample. However, at non-zero temperature one also has to satisfy
the condition $\tau_k\ll\hbar/T$, the time of phase breaking.
So, the temperature interval where our estimate of the noise
holds is $T\le E_c$.


\section{Conclusions}
We introduced a quantum-mechanical scheme that gives complete
statistical description of electron transport. It involves a
spin $1/2$ coupled to the current so that the spin precession
measures transmitted charge. The off-diagonal part of the spin
density matrix, taken as a function of the coupling constant,
gives the generating function for the electron counting statistics.
We find the statistics in a single-channel ideal conductor for
arbitrary relation between temperature and voltage. In
equilibrium, the counting statistics are gaussian, both for
zero-point fluctuations and at finite temperature. At constant
voltage and low temperature the statistics are Bernoullian and
the distribution is binomial.

The theory leads to interesting conclusions applied to the
current fluctuations produced by a voltage pulse. In this case,
the noise has phase sensitivity: it oscillates as function of
Faraday's flux, $c\int V(t)dt$, reaching minimum at integer
fluxes. We studied the noise as function of the shape of the
voltage pulse and found optimal time dependence that provides
absolute minimum of the noise for given average transmitted
charge. Solution displays interesting analogy with the problem
of instantons in the field theories obeying duality symmetry. Optimal
time dependence is a sum of Lorentzian peaks of voltage, each
corresponing to a soliton of flux. The change of flux for a
soliton is equal to the flux quantum $\Phi_0$. The solitons are
interpreted in terms of the binomial statistics picture of
charge fluctuations as attempts to transmit electrons, one
electron per soliton.


\appendix
\section{Larmor clock measurement of tunneling time}
How long does it take a particle to tunnel under a barrier?
More precisely, suppose a particle of energy $E$ is moving in
one dimension, and is scatterred on a potential barrier:
  \be
i{\partial\over\partial t}\psi (x,t)= \left[-{1\over 2}
{\partial^2\over\partial x^2}+U(x)\right]\psi (x,t)\ .
   \ee
What is the probability that during the scattering the particle
spends time $\tau$ within the region $a<x<b$ under the barrier?
Questions of that kind arise naturally in discussion of any
quantum-mechanical process that takes finite time, like nuclear
or chemical reactions, resonance scattering, or tunneling.

There   have   been   several    attempts    to    treat    such
problems\cite{L-clock}  that  resulted  in  formulation of a very
interesting concept of Larmor clock. It  has  various  analogies
with  the spin galvanometer discussed above, and it seems useful
to review the Larmor clock here  using  the  same  language.  The
Larmor  clock  uses  an  auxiliary  spin  $1/2$  attached to the
scattering particle, and an auxiliary  constant  magnetic  field
$\omega$ localized within the region of interest, $a<x<b$,
  \be\label{Linter}
\widehat{\cal H}_{int}= -{1\over2} \omega\sigma_z\int\limits^b_a
\psi^+(x)\psi(x)dx \ .
  \ee
The choice of coupling is such that the spin  precession angle
is proportional to the time spent in the region $a<x<b$.
The difference from our spin-galvanometer is that the spin is
not stationary, but travels with the particle, and also that the spin
is coupled to the particle density, rather than to the current.

To find the distribution of times one has to write down the
system density matrix evolved in time, and take partial trace
over the particle outgoing states. (We assume that one does not
have to distinguish between different results of scattering, and
is interested in the tunneling time only, regardless of whether
the particle went through the barrier, or has been reflected.)
Then, by following the argument of Sec.~III one obtains the spin density
matrix:
  \begin{equation} \label{Lmatr}
{\widehat {\rho}}_s(t)=\left[\begin{array}{cc}
  {\widehat {\rho}}_{\uparrow \uparrow}(0) &
  {\chi}(\omega){\widehat {\rho}}_{\uparrow \downarrow}(0) \\
  {\chi}(-\omega){\widehat {\rho}}_{\downarrow \uparrow}(0) &
  {\widehat {\rho}}_{\downarrow \downarrow}(0)\\
\end{array}
\right] .
  \end{equation}
Here
  \be
\chi(\omega)={\rm tr}_e( e^{-i\widehat{\cal H}_{\omega}t} {\widehat {\rho}_e}
e^{i\widehat{\cal H}_{-\omega}t} )\ ,
  \ee
where $e^{-i\widehat{\cal H}_{\omega}t}$ is the evolution operator for
the one-particle problem with no spin:
  \be\label{Lhamilt}
i{\partial\over\partial t}\psi (x,t)=\left[-{1\over 2}
{\partial^2\over\partial x^2}+U(x)-{1\over2}\omega \theta_{ab}(x)
\right]\psi (x,t)\ ,
   \ee
where $\theta_{ab}=\theta(x-a)\theta(b-x)$.
The auxiliary magnetic field $\omega$ now turns into a constant
potential within the region $a<x<b$. Here again, with the spin
degrees of freedom taken care of by (\ref{Lmatr}), we are left
with a single particle problem. By using cyclic property of the
trace one finds
  \begin{equation}\label{Lchi}
\chi(\omega)=\langle e^{i\widehat{\cal H}_{-\omega}t}
e^{-i\widehat{\cal H}_{\omega}t}\ \rangle\ .
   \end{equation}
Here the brackets $\langle...\rangle$  mean  averaging  over
the  particle  initial state. Note that $\chi(\omega)$ is written in
terms of a purely single particle problem,  not  involving  spin
variables.

The  quantity $\chi(\omega)$ obtained by measuring precession of the
spin is a generating function for  the  distribution  of  times,
which is clear from the Fourier transform
  \be\label{Lgener}
\chi(\omega)=\int P(\tau)e^{i\omega\tau}d\tau\ .
  \ee
The  probabilities  $P(\tau)$  of different precession angles of
the  spin  should  be  interpreted  as   the   scattering   time
distribution.

The probabilities $P(\tau)$ defined by (\ref{Lhamilt}),
(\ref{Lchi}), and (\ref{Lgener}) have several interesting
properties:\\
  {\it a)} $\int P(\tau)d\tau=1$;\\
  {\it b)} $P(\tau)$ are real numbers;\\
  {\it c)} $P(\tau)$ vanish at negative times $\tau<0$.\\
The normalization property {\it a)} is derived from (\ref{Lchi})
by setting $\omega=0$. Property {\it b)} (real valuedness)
is derived from $\chi(-\omega)=\bar\chi(\omega)$
which follows from (\ref{Lchi}). The causality property {\it c)}
follows from considering the evolution in the problem
(\ref{Lhamilt}) with $\omega$ continued to complex values. One
notes that both the solution $\psi(x,t)$ of Eq.~(\ref{Lhamilt})
and the evolution operator $e^{-i\widehat{\cal H}_{\omega}t}$ are
regular in the upper half-plane ${\rm Im}\ \omega>0$, which means
that the same is true for $\chi(\omega)$. From that, the
causality property {\it c)} follows by the usual argument using
Cauchy theorem in the integral
  \be\label{Lprob}
P(\tau)=\int^{\infty}_{-\infty}
\chi(\omega)e^{-i\omega\tau}{d\omega\over2\pi}
  \ee
by closing the integration contour in the upper half-plane.

The properties {\it a), b)} and {\it c)} suggest that $P(\tau)$, so far
defined formally as Fourier spectrum of $\chi(\omega)$, can have
a meaning of probability. However, generally the sign of
$P(\tau)$ can be either positive or negative, which makes the
probabilistic interpretation problematic.

For the one  particle  problem  one  can  write  the  generating
function  $\chi(\omega)$  in  terms of the scattering amplitudes $A$
and  $B$.  For  that, it is convenient to  use   the   expressions
(\ref{A17}),  (\ref{A18}) for the evolution operator in terms of
the   scattering   matrix  $\widehat{\cal  S}$,  written  using  the
wave-packet scattering states  (\ref{A14}).  Specializing  to  one
particle and taking partial trace, one finds
  \be
\chi(\omega)=\bar A_{-\omega}(E) A_{\omega}(E)
+\bar B_{-\omega}(E)B_{\omega}(E)\ ,
  \ee
where  $A(\omega)$  and  $B(\omega)$ are the transmission and reflection
amplitudes of the problem (\ref{Lhamilt}) taken at the energy $E$ of
incident particle.

To see the Larmor clock working, let us consider an example of
resonance scattering, where a particle is scattered on a
potential forming a quasibound state of life-time $\Gamma$.
Using the method described above one can find the distribution
of times it takes the particle to scatter. For simplicity,
suppose that the particle can be only reflected, but not
transmitted ($A=0$). Then the reflection amplitude as function
of energy is given by the Breit-Wigner formula:
  \be
B(E)={E-E_0-i\Gamma/2\over E-E_0+i\Gamma/2}\ .
  \ee
Turning on the field $\omega$ in the quasibound state region is
equivalent to shifting the resonance energy: $E_0\to E_0-\omega/2$.
Thus, the generating function of the time distribution is
  \be
\chi(\omega)={\varepsilon  -\omega+i\Gamma\over \varepsilon  -\omega-i\Gamma}
{\varepsilon  +\omega-i\Gamma\over \varepsilon  +\omega+i\Gamma}\ ,
  \ee
where $\varepsilon  =2(E-E_0)$. The distribution $P(\tau)$ is found by
Fourier transform:
  \ber
P(\tau)&=&\int\chi(\omega)e^{-i\omega \tau}{d\omega\over2\pi}\cr
&=&\delta(\tau)-{4\Gamma\over\varepsilon }\left(
\Gamma \sin \varepsilon \tau- \varepsilon  \cos \varepsilon \tau \right)
e^{-\Gamma\tau}\cr
&=&{\partial\over\partial\tau}\left(\theta(\tau)-{4\Gamma\over\varepsilon }
\sin \varepsilon \tau e^{-\Gamma\tau}\right)\ .
  \eer
The $\delta$-function term corresponds to the non-resonance
scattering channel. Other terms describe dwelling in the
quasibound state. In this example $P(\tau)$ is changing sign,
which makes the  probabilistic interpretation ambiguous.

The paradox arising due to negative $P(\tau)$ is only an
apparent one. Really, the measurement of time performed by the
Larmor clock is not the usual quantum-mechanical measurement,
since the time is not an operator, and thus it cannot be
measured in the same sense as other quantum-mechanical
observables. This should be contrasted with the measurement of
charge described above. Although the spin precession measurement
scheme we use looks quite similar to the Larmor clock, there is
a difference: Electric charge is an observable in the usual
quantum-mechanical sense, it takes quantized integer values, and
the probabilities of those values resulting from our calculation
are non-negative.

\section{Bosonization calculation of counting statistics}

In order to find generating function of counting statistics for
a single channel conductor, we have to evaluate
  \be\label{A-chi}
\chi(\lambda)=\langle \exp i\lambda \widehat N_t\rangle\ ,
  \ee
where $\widehat N_t=\int
(c^+_{1,\tau}c_{1,\tau}c^+_{2,\tau}c_{2,\tau}) d\tau$, and
$c_{i,\tau}$, $c^+_{i,\tau}$ are canonical Fermi operators.

In one dimension, there is an equivalence between ideal Fermi
gas and harmonic Bose chain, which provides a representation of
the Fermi problem in terms of free bosons, known as the
bosonization transformation\cite{Mahan,ortho,Haldane}. This
representation facilitates calculting averages like
(\ref{A-chi}), since they are being transformed to the form of a
gaussian average\cite{ortho}.

According to the bosonization theory, bosonic Hamiltonian
representing the fermionic problem is written as
   \be\label{Bhamilt}
\widehat{\cal H}_{\rm Bose}={\hbar v_F\over 4\pi}\int
:(\nabla\theta_L)^2:+ :(\nabla\theta_R)^2: dx\ ,
  \ee
where $\theta_{L(R)}(x)$ are Bose operators,
   \be\label{A-commutation}
[\nabla\theta_{L(R)}(x),\theta_{L(R)}(y)]=\pm 2\pi i\delta(x-y)\ .
   \ee
Connection to the fermionic problem is given as a relation
between the densities of the left- and right-moving fermions,
$\widehat\rho_i(x)= c^+_{i,x}c_{i,x}$, $i=1,2$, and the bosonic
variables $\theta_{L(R)}(x)$, written as
  \be
\widehat\rho_{1(2)}(x) ={1\over2\pi} \nabla\theta_{L(R)}(x)\ .
  \ee
One notes that the operator $\widehat N_t$ in (\ref{A-chi}) is
linear in the densities $\widehat\rho_i$, and thus it is
represented by an expression linear in the bosonic variables,
  \be
\widehat N_t={1\over2\pi}
\Bigl(\theta_{L}(t)- \theta_{L}(0)- \theta_{R}(t)+ \theta_{R}(0)\Bigr)\ ,
  \ee
which turns the average in (\ref{A-chi}) into a gaussian type.

Therefore, the  average of (\ref{A21}) is equal to the product of averages
   \be\label{A-chi-prod}
\chi(\lambda)=
\langle \exp{i\lambda\over2\pi} (
\theta_{L}(t)- \theta_{L}(0) ) \rangle
\langle \exp{-i\lambda\over2\pi} (
\theta_{R}(t)- \theta_{R}(0) )\rangle
   \ee
taken over the ground state of the Hamiltonian (\ref{Bhamilt}).
To perform the average in (\ref{A-chi-prod}), it is sufficient
to deal with the average over $\theta_L$'s, because of the
left-right symmetry of the problem.

Let us write $\theta_L(x)$ in terms of bosonic operators of plane waves:
  \ber
\theta_L(x)&=&\sum\limits_{k>0}\left({2\pi\over k}\right)^{1/2}\Bigl[
e^{ikx}b_k+e^{-ikx}b^+_k\Bigr]\ ;\cr
\nabla\theta_L(x)&=&\sum\limits_{k>0} (2\pi k)^{1/2} i  \Bigl[
e^{ikx}b_k-e^{-ikx}b^+_k\Bigr]\ .
  \eer
One checks that the commutation relations (\ref{A-commutation})
are consistent with canonical commutation relations between
$b_k$ and $b^+_{k'}$. (The Hamiltonian of left-moving fermions
is represented by $\widehat{\cal H}_L=\sum_{k>0}vkb^+_kb_k$.)
The quantity $\theta_L(t)-\theta_L(0)$ appearing in the average
(\ref{A-chi-prod}) is written as
  \be
\sum\limits_{k>0}\left({2\pi\over k}\right)^{1/2}\Bigl[
(e^{ikvt}-1)b_k+ (e^{-ikvt}-1)b^+_k\Bigr]\ .
  \ee
We evaluate the average
  \ber
& &\langle\!\langle (\theta_L(t)-\theta_L(0))^2\rangle\!\rangle=
\sum\limits_{k>0}{2\pi\over k}|e^{ikvt}-1|^2(2N_{\rm Bose}(kv/T)+1) \cr
& &=4\int\limits_{-\infty}^{\infty}
{dk\over|k|}\sin^2(vkt/2)\coth(vk/2T)=
2\ln\left({1\over\pi T\delta}\sinh(\pi T t)\right)\ .
  \eer
This expression equals $(2\pi)^2$ times
the function $f(t,T)$ computed in
(\ref{compute-f}). From that, we find the average
(\ref{A-chi-prod}) to be
  \ber
& &\chi(\lambda)=
\langle \exp{i\lambda\over2\pi} (
\theta_L(t)-\theta_L(0))\rangle^2 \cr
& &=\exp\Bigl[-\Bigl({\lambda\over2\pi}\Bigr)^2
\langle\!\langle (\theta_L(t)-\theta_L(0))^2\rangle\!\rangle\Bigr]
=\exp\Bigl(-\lambda^2f(t,T)\Bigr)\ ,
  \eer
which is the desired result.

Periodicity of $\chi(\lambda)$ in $\lambda$, corresponding to
the charge quantization, is recovered if one corrects the
relation between $\widehat\rho_i(x)$ and $\theta_i(x)$, in order
to take into account the integer-valuedness of the particle
number $\widehat N_t$. Using the relation,
$\widehat\rho_i(x)=\sum\limits_n e^{in\theta(x)}$, and
performing the average, one arrives at the result (\ref{A22}).


\begin{references}
\bibitem{review}
Mesoscopic Phenomena in Solids,
B.~L.~Altshuler, P.~A.~Lee, and R.~A.~Webb, eds., (North Holland, 1991)
\bibitem{9}
R.~Landauer,  in: Localization,  Interaction   and
Transport Phenomena, eds. B.~Kramer, G.~Bergmann and Y.~Bruynsraede
(Springer, Heidelberg, 1985) Vol.{\bf 61}, 38;\\
R.~Landauer, in: W.~van~Haeringen  and
D.~Lenstra  (eds.),  Analogies  in  Optics and Micro Electronics,
243-257, Kluwer Academic Publishers (1990);\\
Y.~Imry, {\it Directions in Condensed Matter Physics}, 101,
G.~Grinstein and G.~Mazenko, eds., (World Scientific, Singapore, 1986)
\bibitem{2}
G. B. Lesovik, JETP Letters {\bf 49}, 594 (1989)
\bibitem{3}
B.~Yurke and G.~P.~Kochanski, Phys. Rev. B{\bf 41}, 8184 (1990)
\bibitem{4}
M.~B{\"u}ttiker, Phys. Rev. Lett. {\bf 65}, 2901 (1990);
Phys. Rev. B{\bf 46}, 12485 (1992)
\bibitem{5}
S.-R.~E.~Yang, Solid State Commun.{\bf 81}, 375 (1992)
\bibitem{6}
C.~W.~J.~Beenakker, M.~B{\"u}ttiker, Phys. Rev. B{\bf46}, 1889 (1992)
\bibitem{7}
Th.~Martin and R.~Landauer, Phys. Rev. B{\bf 45}, 1742 (1992)
\bibitem{deJong}
M.~J.~M.~de~Jong and C.~W.~J.~Beenakker, Phys. Rev. B{\bf 46}, 13400 (1992)
\bibitem{LL}
L.~S.~Levitov and G.~B.~Lesovik,
JETP Letters {\bf 58} (3), 230-235 (1993)
\bibitem{1'94}
L.~S.~Levitov and G.~B.~Lesovik,
Quantum Measurement in Electric Circuit,
preprint cond-mat/9401004
\bibitem{2'93}
D.~A.~Ivanov and  L.~S.~Levitov,
JETP Letters {\bf 58}(6), 461 (1993)
\bibitem{C9}
H.-W.~Lee and  L.~S.~Levitov, Orthogonality catastrophe in a  mesoscopic
conductor due to a time-dependent flux, preprint cond-mat/9312013
\bibitem{2'94}
B.~L.~Altshuler, L.~S.~Levitov, and  A.~Yu.~Yakovets,
JETP Letters {\bf 59},  857 (1994) (in Russian: Pis'ma v ZhETF,
vol. {\bf59},  821 (1994))
\bibitem{3'94}
L.~S.~Levitov and G.~B.~Lesovik,
Phys. Rev. Lett. {\bf 72},  538 (1994)
\bibitem{3'95}
D.~A.~Ivanov, H.-W.~Lee, and L.~S.~Levitov,
Coherent states of alternating current,
preprint cond-mat/9501040,
to appear in  Phys. Rev. {\bf B}
\bibitem{1'95}
H.-W.~Lee and L.~S.~Levitov,
Estimate of minimal noise in a quantum conductor,
preprint cond-mat/9507011
\bibitem{MandelGlauber}
R.~J.~Glauber, Phys. Rev. Lett. {\bf 10}, 84
(1963); Phys. Rev. {\bf 130}, 2529 (1963);\\
L.~Mandel and E.~Wolf, Rev. Mod. Phys. {\bf 37}, 231 (1965)
\bibitem{Gradinger} C.~W.~Gardinger, Quantum Noise, Chapter 8,
Springer-Verlag (1991);\\
J.~R.~Klauder and E.~C.~G.~Sudarshan, Fundamentals of Quantum Optics,
Chapter 8, W.A.Benjamin, Inc., N.Y. (1968)
\bibitem{0-noise} R.~H.~Koch, D.~van~Harlingen, and J.~Clarke,
Phys. Rev. B{\bf 26}, 74 (1982)
\bibitem{e-count} S.~Saito, {\it et al.}, Phys. Lett. A{\bf 162}, 442 (1992)
\bibitem{Leggett} A.~J.~Leggett, Progr. Theor. Phys. Suppl.{\bf 69}, 80 (1980)
\bibitem{L-clock} M.~ B\"uttiker, Phys. Rev. B{\bf 27}, 6178 (1983);\\
A.~I.~Baz', Sov. Phys. JETP {\bf 20}, 1261 (1965);\\
A.~I.~Baz', Ya.~B.~Zeldovich and A.~M.~Perelomov,  Scattering,
Reactions and Decay in Nonrelativistic Quantum Mechanics, Israel
Program for Scientific Translations, Jerusalem (1969)
\bibitem{Mahan}
G. D. Mahan, Many Particle Physics, Secs.~4.4,~8.3,
(2-nd edition, Plenum Press, 1990)
\bibitem{ortho}
K.~D.~Schotte and U.~Schotte, Phys. Rev. {\bf 182}, 479 (1969);
\bibitem{Haldane} F. D. M. Haldane, J. Phys. C{\bf 14}, 2585-2609 (1981)
\bibitem{Anderson}
P.~W.~Anderson, Phys. Rev. Lett. {\bf 18}, 1049 (1967)
\bibitem{Nozieres}
P. Nozi\`eres, C. T. deDominicis, Phys. Rev. {\bf 178}, 1084 (1969)
\bibitem{C10}
A.~A.~Belavin  and  A.~M.~Polyakov,  Pis'ma
ZhETF\ {\bf 22}, 503 (1975);\\
see  also:  A.~M.~Polyakov,  Gauge  Fields   and
Strings, Chap.6, Sect.1 (Harwood Academic Publishers, 1987)
\end{references}
\end{document}